\documentclass[preprint,aps]{revtex4}
\usepackage{graphicx,slashed,color}

\begin{document}
\draft
\title{Parameterization of General $Z$-$\gamma$-$Z'$ Mixings in an Electroweak Chiral Theory}

\author{Ying Zhang$^1$\footnote{{\it Email address}: hepzhy@mail.xjtu.edu.cn},
Qing Wang$^{2,3}$\footnote{Corresponding author at: Department of Physics, Tsinghua University, Beijing 100084,P.R.China\\
{\it Email address}:~wangq@mail.tsinghua.edu.cn}}

\address{$^1$School of Science, Xi'an Jiaotong University, Xi'an, 710049, P.R.China\\
    $^2$Center for High Energy Physics, Tsinghua University, Beijing 100084, P.R.china\\
    $^3$Department of Physics,Tsinghua University,Beijing 100084,P.R.China}
\date{Mar 1, 2011}

\begin{abstract}
A new general parameterization with eight mixing parameters  among  $Z$, $\gamma$ and an extra neutral gauge boson $Z'$ is proposed and subjected to phenomenological analysis. We show that in addition to the conventional Weinberg angle $\theta_W$, there are seven other phenomenological parameters $G'$, $\xi$, $\eta$, $\theta_l$, $\theta_r$, $r$, $l$ for the most general $Z$-$\gamma$-$Z'$ mixings, in which parameter $G'$ arises due to the presence of an extra Stueckelberg-type mass coupling. Combined with the conventional $Z$-$Z'$ mass mixing angle $\theta'$, the remaining six parameters $\xi$, $\eta$, $\theta_l$-$\theta'$, $\theta_r$-$\theta'$, $r$, $l$ are caused by general kinetic mixings. In all the eight phenomenological parameters $\theta_W$, $G'$, $\xi$, $\eta$, $\theta_l$, $\theta_r$, $r$, $l$, we can determine the $Z$-$Z'$ mass mixing angle $\theta'$ and the mass ratio $M_{Z}/M_{Z'}$. The $Z$-$\gamma$-$Z'$ mixings we discuss are based on the model-independent description of the extended electroweak chiral Lagrangian (EWCL) previous proposed by us. In addition, we show that there are eight corresponding independent theoretical coefficients in our EWCL which are fully fixed by our eight phenomenological mixing parameters. We further find that the experimental measurability of these eight parameters does not rely on the extended neutral current for $Z'$, but depends on the $Z-Z'$ mass ratio.

\end{abstract}

\maketitle

\section{Introduction}

One of the simplest and more popular gauge extensions of the standard model (SM) is to add an extra $U(1)$ group associated with the $Z'$ gauge boson to the
electroweak gauge group $SU(2)_L\otimes U(1)_Y$, that constitutes one of the "hot spots" in high energy physics today. The extra gauge boson $Z'$ is the carrier of a new gauge force corresponding to the smallest gauge group extensions that plays a crucial role in cosmology, GUT, SUSY and various strong coupling new physics theories associated with new physics beyond SM (for the latest review see \cite{LangackerRMP2008}). As long as there exists a $Z'$ particle, it will shift observables from present physics by mixing with the standard electroweak neutral gauge bosons, $\gamma$ and $Z$. The corrections, however, depend on details of the model set-up, and especially on the way the neutral gauge bosons mix. A model-independent way to figure out these mixings is through phenomenological requirements and constraints. Usually, theorists only consider minimal $Z$-$Z'$ mass mixing \cite{Langacker2009}. A massless photon constrains any possible extension of the mass mixings matrix to be of Stueckelberg-type \cite{Zhang2008JHEP}.   However, theory and phenomenology do not forbid  general three-body $Z$-$\gamma$-$Z'$ kinetic mixing.  In the literature only a few examples have been considered, such as, the special kinetic mixings given in \cite{Holdom} and \cite{Daniel}. A general model-independent description of $Z$-$\gamma$-$Z'$ mixing is needed to enable data analysis and experimental searches for $Z'$ to be more specific and effective, particularly in light of the progress made in the LHC and Tevatron experiments. With this motivation, we are prompted to study the most general gauge boson mixing. In fact, a general description of the $Z'$ interaction with SM particles has already been given in our previous work \cite{Zhang2008JHEP,Zhang2009JHEP} in which $Z'$ is  regarded as a gauge boson of a broken $U(1)'$ symmetry and the conventional EWCL is extended to include this extra broken $U(1)'$ symmetry from original $SU(2)_L\otimes U(1)_Y\rightarrow U(1)_{em}$ to $SU(2)_L\otimes U(1)_Y\otimes U(1)'\rightarrow U(1)_{em}$. In Ref.\cite{Zhang2008JHEP}, the bosonic part up to order $p^4$ of the most general EWCL involving this $Z'$ boson and discovered particles has been proposed that describes the most general $Z$-$\gamma$-$Z'$ mixings. In Ref.\cite{Zhang2009JHEP}, various $Z$-$\gamma$-$Z'$ mixings that have appeared in the literature are shown to be included in our EWCL formalism and are further classified into five simple groupings. However, the expressions given in  \cite{Zhang2008JHEP,Zhang2009JHEP} for these $Z$-$\gamma$-$Z'$ mixings are complex and are not suitable for phenomenological investigations.

It is the purpose of this paper to improve this shortcoming by setting up a more general parameterization for all $Z$-$\gamma$-$Z'$ mixings to facilitate present and future phenomenological analysis in the EWCL given by \cite{Zhang2008JHEP}. We will discuss the physical meaning, origin and experimental measurability of these parameters within new parameterization. We show that there are eight independent degree of freedoms and all complexities of the mixing can be absorbed into eight phenomenological parameters $\theta_W$, $G'$, $\xi$, $\eta$, $\theta_l$, $\theta_r$, $r$, $l$, for which all but the  traditional Weinberg mixing angle $\theta_W$ and the Stueckelberg-type coupling $G'$, combine with the conventional $Z$-$Z'$ mass mixing angle $\theta'$, and the remaining six parameters $\xi$, $\eta$, $\theta_l$-$\theta'$, $\theta_r$-$\theta'$, $r$, $l$ are caused by general kinetic mixings. We will explicitly construct quantitative relations among these mixing parameters and those related to theoretical coefficients appearing in the underlying EWCL.

This paper is organized as follows.  In Section II, we give a short review of the relevant parts associated with the $Z$-$\gamma$-$Z'$ kinetic and mass mixings  from the EWCL given in Ref.\cite{Zhang2008JHEP}, and introduce the mixing matrix. In Section III, we explain the physical meaning and origin of the eight parameters describing the mixing matrix by diagonalizing the mass-squared and kinetic matrices, and construct the relations among the various mixing matrix elements and coefficients in our EWCL. In Section IV, we first discuss the experimental measurability of parameters arising in our new parameterization,  then express the EWCL coefficients related to  $Z$-$\gamma$-$Z'$ mixings in these eight parameters that transfers the  measurability from the mixing parameters to the relevant EWCL coefficients. Section V presents a summary.
\section{Review of the kinetic and mass mixings from EWCL}

 We begin the discussion by first reviewing the EWCL of $Z'$ established in \cite{Zhang2008JHEP}. The general Lagrangian describing the gauge symmetry breaking $SU(2)_L\otimes U(1)_Y\otimes U(1)'\rightarrow U(1)_{em}$ independent of the details of the symmetry breaking can be constructed in terms of $2\times2$ non-linear Goldstone field $\hat{U}$ with the following covariant derivative
\begin{eqnarray*}
D_\mu \hat{U}=\partial_\mu\hat{U}+igW_\mu\hat{U}-i\hat{U}(g'\frac{\tau_3}{2}+\tilde{g}')B_\mu-i{g''}\hat{U}X_\mu\;,
\end{eqnarray*}
where $W_\mu$, $B_\mu$ and $X_\mu$ are gauge bosons corresponding to $SU(2)_L$, $U(1)_Y$ and $U(1)'$, respectively.
Here, carets are used to distinguish extended $U(1)'$ breaking quantities from the traditional electroweak breaking quantities in \cite{AppelquistPRD1993}. $g$, $g'$, $g''$ and $\tilde{g}'$ are $SU(2)_L$ coupling, conventional $U(1)_Y$ coupling,
$U(1)'$ coupling and special Stueckelberg-type gauge coupling, respectively.

In paper \cite{Zhang2008JHEP}, the bosonic part  of the Lagrangian up to order $p^4$  has been presented.
Because of  our interests here in $Z'$ mixing effects, we focus only on the
neutral gauge boson mixing parts, which can be divided into a mass part $\mathcal{L}_{M}$
\begin{eqnarray}
\mathcal{L}_M&=&-\frac{1}{4}f^2\mathrm{tr}[\hat{V}_\mu^2]
+\frac{1}{4}\beta_1f^2\left(\mathrm{tr}[T\hat{V}_\mu]\right)^2
+\frac{1}{4}\beta_2f^2\mathrm{tr}[\hat{V}_\mu]\mathrm{tr}[T\hat{V}^\mu]
+\frac{1}{4}\beta_3f^2\left(\mathrm{tr}[\hat{V}_\mu]\right)^2\nonumber\\
&&\hspace*{-2.1cm}\stackrel{\mbox{\tiny unitary gauge}}{======}\frac{f^2}{8}(1\!-\!2\beta_1)(gW^3_\mu\!-g'B_\mu)^2
+\frac{f^2}{2}(1\!-\!2\beta_3)(g^{\prime\prime}X_\mu\!+\tilde{g}'B_\mu)^2\nonumber\\
    &&+\frac{f^2}{2}\beta_2(g^{\prime\prime}X_\mu+\tilde{g}'B_\mu)(gW^{3,\mu}-g'B^\mu)
=\frac{1}{2}\mathcal{V}^T_\mu\mathcal{M}^2_0\mathcal{V}_\mu
\end{eqnarray}
and kinetic part $\mathcal{L}_K$
\begin{eqnarray}
\mathcal{L}_K
&=&-\frac{1}{4}B_{\mu\nu}^2
    -\frac{1}{2}\mathrm{tr}[W_{\mu\nu}^2]
    -\frac{1}{4}X_{\mu\nu}^2
+\frac{1}{2}\alpha_1gg'B_{\mu\nu}\mathrm{tr}[TW^{\mu\nu}]
+\frac{1}{4}\alpha_8g^2\left(\mathrm{tr}[TW_{\mu\nu}]\right)^2
\nonumber\\
&&+g{g''}\alpha_{24}X_{\mu\nu}\mathrm{tr}[TW^{\mu\nu}]
+g'{g''}\alpha_{25}B_{\mu\nu}X^{\mu\nu}\nonumber\\
&&\hspace*{-2.1cm}\stackrel{\mbox{\tiny unitary gauge}}{======}-\frac{1}{4}B_{\mu\nu}B_{\mu\nu}
    -\frac{1}{4}X_{\mu\nu}X^{\mu\nu}
     -\frac{1}{4}(1\!-\!\alpha_8g^2)(\partial_\mu W^3_\nu\!-\partial_\nu W^3_\mu)^2\nonumber\\
   &&+\frac{1}{2}\alpha_1gg'B_{\mu\nu}(\partial_\mu W^3_\nu\!-\partial_\nu
    W^3_\mu)+gg^{\prime\prime}\alpha_{24}X^{\mu\nu}(\partial_\mu W^3_\nu-\partial_\nu W^3_\mu)
    +g'g^{\prime\prime}\alpha_{25}B_{\mu\nu}X^{\mu\nu}\nonumber\\
&=&-\frac{1}{4}\mathcal{V}^T_{\mu\nu}\mathcal{K}_0\mathcal{V}^{\mu\nu}.
\end{eqnarray}
Here, $T\equiv\hat{U}^\dag\tau_3\hat{U}$ and $\hat{V}_\mu\equiv(\hat{D}_\mu \hat{U}) \hat{U}^\dag$ are $SU(2)_L$ covariant operators.
In $\mathcal{L}_M$, the first term is the conventional non-linear $\sigma$ model term and the fourth term is a new  non-linear $\sigma$ model term due to the presence of the $U(1)'$ Goldstone boson. The  second term is the conventional custodial symmetry breaking term. The third term is the
mixing of the second and fourth terms. For $\mathcal{L}_K$, with the exception of the standard kinetic terms for the $U(1)_Y$, $SU(2)_L$ and $U(1)'$ gauge bosons, the terms with coefficients $\alpha_1$, $\alpha_{24}$ and $\alpha_{25}$ are the kinetic mixing terms between $U(1)$ and the  diagonal part of the $SU(2)_L$ gauge fields,  between $U(1)'$ and the diagonal part of the $SU(2)_L$ gauge fields,  and  between $U(1)$ and $U(1)'$ gauge fields, respectively. The term with coefficients $\alpha_8$  is the correction term for the diagonal part of $SU(2)_L$ gauge field. These coefficients parameterize the most general kinetic mixing among the $Z$-$\gamma$-$Z'$ bosons. For convenience, all these terms have been abbreviated into matrix forms in the unitary gauge $\hat{U}=1$ in the gauge boson vector $\mathcal{V}^T_\mu=(W^3_\mu,B_\mu,X_\mu)$,
the field strength tensor $\mathcal{V}_{\mu\nu}\equiv\partial_\mu\mathcal{V}_\nu-\partial_\nu\mathcal{V}_\mu$, the mass-squared matrix $\mathcal{M}_0^2$
and the kinetic matrix $\mathcal{K}_0$. The mass-squared and kinetic matrices are
\begin{eqnarray}
&&\hspace*{-1cm}\mathcal{M}_0^2=f^2\left(\begin{array}{ccc}
\frac{g^2}{4}(1\!-\!2\beta_1)&-\frac{gg'}{4}(1\!-\!2\beta_1)+\frac{g\tilde{g}'}{2}\beta_2
&\frac{gg''}{2}\beta_2\\
-\frac{gg'}{4}(1\!-\!2\beta_1)+\frac{g\tilde{g}'}{2}\beta_2&\frac{g'^2}{4}(1\!-\!2\beta_1)+\tilde{g}'^2(1\!-\!2\beta_3)-g\tilde{g}'\beta_2
&-\frac{g'g''}{2}\beta_2+g''\tilde{g}'(1\!-\!2\beta_3)\\
\frac{gg''}{2}\beta_2&-\frac{g'g''}{2}\beta_2+g''\tilde{g}'(1\!-\!2\beta_3)
&g''^2(1\!-\!2\beta_3)\end{array}\right)\;,~~~\\
&&\hspace*{4cm}\mathcal{K}_0=-\frac{1}{4}\left(\begin{array}{ccc}
		1-g^2\alpha_8&-gg'\alpha_1
&-2gg''\alpha_{24}\\
		-gg'\alpha_1&1&-2g'g''\alpha_{25}
		\\
-2gg''\alpha_{24}&-2g'g''\alpha_{25}
&1			\end{array}\right)\;.
\end{eqnarray}
From $\mathcal{M}_0^2$ and $\mathcal{K}_0$, we see that three body $Z$-$\gamma$-$Z'$ mixing is controlled by 11 dimensionless coefficients: 4 gauge couplings $g,g',\tilde{g}',g''$, 3 mass-mixing low-energy constants $\beta_1$, $\beta_2$, $\beta_3$ and 4 kinetic-mixing low-energy constants $\alpha_1$, $\alpha_8$, $\alpha_{24}$, $\alpha_{25}$. Among these, only nine play roles in the sense that we can redefine nine new coefficients by absorbing $\beta_1$ and $\beta_3$ as follows
\begin{eqnarray}
&&\hspace*{-1cm}g'=\frac{\bar{g}'}{\sqrt{1-2\beta_1}}
\hspace*{1.5cm}g=\frac{\bar{g}}{\sqrt{1-2\beta_1}}
\hspace*{1.5cm}g''=\frac{\bar{g}''}{\sqrt{1-2\beta_3}}
\hspace*{1.5cm}{\tilde{g}'}=\frac{\bar{\tilde{g}}'}{\sqrt{1-2\beta_3}}\;,\\
&&\hspace*{-1cm}\beta_2=\bar{\beta}_2\sqrt{1\!-\!2\beta_1}\sqrt{1\!-\!2\beta_3}
\hspace*{0.6cm}\alpha_a=gg'\bar{\alpha}_1\hspace*{0.6cm}\alpha_b=g^2\bar{\alpha}_8
\hspace*{0.6cm}\alpha_{c}=gg''\bar{\alpha}_{24}
\hspace*{0.6cm}\alpha_{d}=g'g''\bar{\alpha}_{25}\;.~~
\end{eqnarray}
Then $\mathcal{M}_0^2$ and $\mathcal{K}_0$ of these redefined nine coefficients become
\begin{eqnarray}
\mathcal{M}_0^2&=&f^2\left(\begin{array}{ccc} \frac{\bar{g}^2}{4}&-\frac{\bar{g}\bar{g}'}{4}+\frac{\bar{g}\bar{\tilde{g}}'}{2}\bar{\beta}_2
&\frac{\bar{g}\bar{g}''}{2}\bar{\beta}_2
\\ -\frac{\bar{g}\bar{g}'}{4}+\frac{\bar{g}\bar{\tilde{g}}'}{2}\bar{\beta}_2&\frac{\bar{g}'^2}{4}+\bar{\tilde{g}}'^2-\bar{g}\bar{\tilde{g}}'\bar{\beta}_2
&-\frac{\bar{g}'\bar{g}''}{2}\bar{\beta}_2+\bar{g}''\bar{\tilde{g}}'\\
	\frac{\bar{g}\bar{g}''}{2}\bar{\beta}_2&-\frac{\bar{g}'\bar{g}''}{2}\bar{\beta}_2+\bar{g}''\bar{\tilde{g}}'
&\bar{g}''^2
	\end{array}\right)\;,\\
\mathcal{K}_0&=&-\frac{1}{4}\left(\begin{array}{ccc}
		1-\alpha_b&-\alpha_a
&-2\alpha_c	\\
		-\alpha_a&1
&-2\alpha_d
		\\
		-2\alpha_c&-2\alpha_d&1		
	\end{array}\right)\;.
\end{eqnarray}
Furthermore there exists a scale symmetry for $\mathcal{M}_0^2$ and $\mathcal{K}_0$, i.e.,  these are invariant under the following transformation determined by an arbitrary parameter $\zeta$,
\begin{eqnarray}
\bar{g}\rightarrow \zeta\bar{g}\hspace*{1cm}\bar{g}'\rightarrow \zeta\bar{g}'\hspace*{1cm}\bar{g}''\rightarrow \zeta\bar{g}''
\hspace*{1cm}\bar{\tilde{g}}'\rightarrow \zeta\bar{\tilde{g}}'\hspace*{1cm}f\rightarrow \frac{1}{\zeta}f
\end{eqnarray}
with $\bar{\beta}_2,\alpha_a,\alpha_b,\alpha_c,\alpha_d$ unchanged. Since the dimensional coefficient $f$ does not enter into the final mixing matrix, the above scale symmetry implies that among the nine redefined theoretical coefficients, only eight of these are independent, and span the largest mixing space for an extra neutral gauge boson $Z'$. We take these eight theoretical coefficients as $\bar{g}/\bar{g}'$, $\bar{\tilde{g}}'/\bar{g}''$, $\bar{g}_Z/\bar{g}''$, $\bar{\beta}_2$, $\alpha_a$, $\alpha_b$, $\alpha_c$, $\alpha_d$  with
\begin{eqnarray}
\bar{g}_Z\equiv\sqrt{\bar{g}^2+\bar{g}'^2}\;.
\end{eqnarray}
These will provide all combinations of extra neutral vector boson corrections to low-energy EW physics via through mixings. As discussed in \cite{Zhang2009JHEP}, then inputting different set of values for
these coefficients, the effective theory can recuperate the  various $Z'$ models that have been presented in the literature. The mixings can be disentangled by diagonalizing the mass-squared matrix $\mathcal{M}_0^2$ and kinetic matrix  $\mathcal{K}_0$ simultaneously, i.e. through introducing in a $3\times 3$ real matrix $U$ which relates the interaction eigenstate $(W^3_\mu,B_\mu,X_\mu)$ to the mass eigenstate  $(Z_\mu,A_\mu,Z'_\mu)$ in the following manner
\begin{eqnarray}
\left(\begin{array}{c}W^3_\mu\\B_\mu\\X_\mu\end{array}\right)=U\left(\begin{array}{c}Z_\mu\\A_\mu\\Z'_\mu\end{array}\right)\label{Uintr}\;.
\end{eqnarray}
The $U$ matrix has to fulfill conditions
\begin{eqnarray}
U^T\mathcal{M}_0^2U=\mathrm{diag}(M_Z^2,0,M_{Z'}^2)\hspace*{4cm}U^T\mathcal{K}_0U=-\frac{1}{4}\mathrm{diag}(1,1,1)\;.\label{diag}
\end{eqnarray}
In Refs.\cite{Zhang2008JHEP,Zhang2009JHEP}, we have already discussed the exact form of $U$, although in practice its physical meaning tends to get lost due to its complex form, and is not suitable in presenting phenomenological arguments. Here, we simplify its expression by re-parameterizing it as follows,
\begin{eqnarray}
&&\hspace{-1cm}U\equiv\left(\begin{array}{ccc}
        s_W\xi+c_Wc_ll ~~&~~s_Wa~~&~~s_W\eta+c_Ws_rr \\
        c_W\xi-s_Wc_ll ~~&~~c_Wa~~
        &~~c_W\eta-s_Ws_rr\\
        -c_W\xi G'+s_Wc_l l G'-s_ll ~~&~~-c_Wa G'~~
        &~~-c_W\eta G'+s_Ws_rr G'+c_rr
    \end{array}\right)=U_0U_1\;,~~\label{Udef}\\
    &&\hspace*{-1cm}U_0\equiv\left(\begin{array}{ccc}
        c_W & s_W & 0\\
        -s_W & c_W & 0\\
        s_W G' & -c_W G' & 1
    \end{array}\right)\hspace*{3cm}U_1\equiv\left(\begin{array}{ccc}
        lc_l & 0 & rs_r
        \\
        \xi & a & \eta
        \\
        -ls_l  & 0& rc_r
    \end{array}\right)\;,
\end{eqnarray}
in which there are three angle parameters $\theta_W,\theta_r,\theta_l$ establishing the trigonometric values $c_i\equiv\cos\theta_i$, $s_i\equiv\sin\theta_i$ for $i=W,l,r$ and
six other mixing parameters $G',a,\xi,\eta,r,l$, totally nine in all. Among these nine parameters, $a=a(\theta_W,\theta_r,\theta_l,G',\xi,\eta,r,l)$ is a single relation determining one of the other eight parameters, the detailed dependence will be given later in (\ref{a}). Thus only eight of nine parameters in (\ref{Udef}) are independent, the degree of freedoms just matches the number of independent theoretical coefficients for electroweak gauge boson mixings that we counted  before. In fact,  because of the massless photon, parameter $a$ is a normalization constant and plays the role of rescaling the photon field, which does not cause observable effects in the two-point vertices involving electroweak gauge bosons. Note that  in the SM tree diagram limit, $U_0$ is a pure Weinberg rotation with $G'=0$ and $U_1$ is a unit matrix with $\theta_l=\theta_r=\xi=\eta=0$ and $l=r=a=1$.

\section{ Phenomenological parameters in terms of diagonalization and EWCL coefficients}

Next, we explain the physical meaning and origin of the eight parameters $\theta_W$, $G'$, $\xi$, $\eta$, $\theta_r$, $\theta_l$, $r$, $l$ by diagonalizing the mass-squared
matrix $\mathcal{M}_0^2$ and kinetic matrix $\mathcal{K}_0$. First, $G'$ is defined in such a way that it relates to the Stueckelberg-type coupling $\bar{\tilde{g}}'$ as
\begin{eqnarray}
G'\equiv\frac{\bar{\tilde{g}}'}{\bar{g}''}=\frac{\tilde{g}'}{g''}\;.\label{Gp}
\end{eqnarray}
i.e.,  $G'$ is derived from the Stueckelberg coupling as the ratio of the Stueckelberg coupling and conventional $U(1)'$ coupling. In our EWCL formalism, the deviation from SM has two sources: a Stueckelberg-type interaction for $B_\mu$ and the extra $U(1)'$ interaction from gauge boson $X_\mu$, with $G'$  the relative ratio of the interaction strengths between these two types of sources. Theoretically $G'$ can take arbitrary real numbers, in particular $G'=\infty$ and $G'=0$ correspond to $g''=0,~\tilde{g}'$ finite and $\tilde{g}'=0,~g''$ finite, respectively. However, phenomenological analysis shows that a very large $G'$ is not physically realistic as Ref.\cite{StSM} gives $G'=\tilde{g}'/g''=1.9/149\approx 0.013$. If we ignore $G'$, the rotation matrix $U_0$ then reverts to the standard Weinberg rotation with Weinberg angle $\theta_W$ defined as
\begin{eqnarray}
c_W\equiv \frac{\bar{g}}{\bar{g}_Z}\hspace*{1cm}s_W\equiv \frac{\bar{g}'}{\bar{g}_Z}\hspace*{2cm}
\mbox{or}\hspace*{2cm}
\theta_W=\arctan\frac{\bar{g}'}{\bar{g}}=\arctan\frac{g'}{g}\;.\label{WeinbergDef}~~~~
\end{eqnarray}
 The Weinberg angle originates from mixing of field $W^{3,\mu}$ and $B^\mu$ and the Weinberg rotation enables the part of the mass matrix associated with $\gamma$ and $Z$ to be diagonalized if the $Z'$ particle and the Stueckelberg coupling are neglected.
Once the Stueckelberg coupling $\bar{\tilde{g}}'$ shows up, there will be off diagonal matrix elements involving
$\gamma$-$Z$ and $\gamma$-$Z'$ mixings. To disentangle these mixings, we add  $G'$ terms to the $U_0$ matrix and after the $U_0$ rotation, we find
\begin{eqnarray}
U_0^T\mathcal{M}_0^2U_0=f^2\left(\begin{array}{ccc} \frac{1}{4}\bar{g}_Z^2&0&\frac{1}{2}\bar{g}_Z\bar{g}''\bar{\beta}_2\\
	0&0&0\\
	\frac{1}{2}\bar{g}_Z\bar{g}''\bar{\beta}_2&0&\bar{g}''^2
	\end{array}\right)\label{MassMix-1}\;.
\end{eqnarray}
This is a typical $Z$-$Z'$ mixing matrix. We apply a further matrix $\tilde{U}_0$ with rotation angle $\theta'$ to diagonalize (\ref{MassMix-1}), i.e.
\begin{eqnarray}
\tilde{U}_0=\left(\begin{array}{ccc}
        c' & 0 & s'\\
        0 & 1 & 0\\
        -s' & 0 & c'
    \end{array}\right)\hspace*{2cm}
\tilde{U}_0^TU_0^T\mathcal{M}_0^2U_0\tilde{U}_0=\mathrm{diag}(M^2,0,{M'}^2)\label{diag1}
\end{eqnarray}
with $c'=\cos\theta',s'=\sin\theta'$. We find that  it fixes the rotation angle $\theta'$ as follows
\begin{eqnarray}
\theta'
=\arctan\frac{\Delta_g-\sqrt{\Delta_g^2+16\bar{g}_Z^2\bar{g}''^2\bar{\beta}_2^2}}{4\bar{\beta}_2\bar{g}''\bar{g}_Z}
\hspace*{2cm}\Delta_g=\bar{g}_Z^2-4\bar{g}''^2\;.\label{thetaprime}
\end{eqnarray}
Hence $\theta'$ originates from the $Z$-$Z'$ mass mixing ,its role being to disentangle this mixing, and appears in most of the new physics models involving the $Z'$ boson.
With the zero eigenvalue in (\ref{diag1}) corresponding to the massless photon, the two other nonzero eigenvalues in (\ref{diag1}) are
\begin{eqnarray}
\frac{M^2}{f^2}=\frac{1}{4}\bar{g}_Z^2c'^2+\bar{g}''^2s'^2-s'c'\bar{g}_Z\bar{g}''\bar{\beta}_2
\hspace*{1cm}
\frac{M'^2}{f^2}=\bar{g}''^2c'^2+\frac{1}{4}\bar{g}_Z^2s'^2+s'c'\bar{g}_Z\bar{g}''\bar{\beta}_2\;.\label{MMprime}
\end{eqnarray}
Here  $M$ and $M'$ are just the $Z$ and $Z'$ masses if there are no Stueckelberg and kinetic mixings. For $g''=\tilde{g}'=0$, (\ref{MassMix-1}) is already diagonal with eigenvalues $\frac{1}{4}f^2\bar{g}_Z^2,0,0$,  and there is no need to apply further rotation; clearly, $\theta'=0$ is given by (\ref{thetaprime}) resulting in a unit matrix $\tilde{U}_0$. This further simplifies the eigenvalues of (\ref{MMprime}) to $M^2/f^2=\frac{1}{4}\bar{g}_Z^2$ and $M^{\prime 2}/f^2=0$. Here $M'=0$ implies the mass of $Z'$ is zero and $Z'$ decouples from $Z$ and $\gamma$.

 After diagonalizing the mass-squared matrix $\mathcal{M}_0^2$, the next logical step is to further diagonalize the kinetic matrix $\mathcal{K}_0$. Considering that after the rotation $U_0\tilde{U}_0$ which diagonalizes $\mathcal{M}_0^2$,  the kinetic matrix $\mathcal{K}_0$ is already transformed to symmetric form
 \begin{eqnarray}
 &&\hspace*{-1.1cm}\tilde{U}_0^TU_0^T\mathcal{K}_0U_0\tilde{U}_0=\left(\begin{array}{ccc}
		k_1&k_2&k_3\\
		k_2&k_4&k_5\\
		k_3&k_5&k_6
	\end{array}\right)
\end{eqnarray}
with
\begin{eqnarray}
k_1&=&1-2s_Ws'c'G'+s_W^2c'^2G'^2+2c_Ws_Wc'^2\alpha_a-c_W^2c'^2\alpha_b\nonumber\\
		&&+(4c_Ws'c'-4c_Ws_Wc'^2G')\alpha_c+(-4s_Ws'c'+4s_W^2c'^2G')\alpha_d\;,\\
k_2&=&c_Ws'G'-c_Ws_Wc'G'^2+(s_W^2-c_W^2)c'\alpha_a-c_Ws_Wc'\alpha_b\nonumber\\
		&&+[2s_Ws'+2(c_W^2-s_W^2)c'G']\alpha_c+(2c_Ws'-4c_Ws_Wc'G')\alpha_d\;,\\
k_3&=&-s_W(s'^2-c'^2)G'+s_W^2s'c'G'^2+2c_Ws_Ws'c'\alpha_a-c_W^2c's'\alpha_b\nonumber\\
		&&+[2c_W(s'^2-c'^2)-4c_Ws_Ws'c'G']\alpha_c+[2s_W(c'^2-s'^2)+4s_W^2s'c'G']\alpha_d\;,\\
k_4&=&1+c_W^2G'^2-2c_Ws_W\alpha_a-s_W^2\alpha_b+4c_Ws_WG'\alpha_c+4c_W^2G'\alpha_d\;,\label{k1}\\
k_5&=&-c_Wc'G'-c_Ws_Ws'G'^2+(s_W^2-c_W^2)s'\alpha_a-c_Ws_Ws'\alpha_b\nonumber\\
		&&-[2s_Wc'-2(c_W^2-s_W^2)s'G']\alpha_c-(2c_Wc'+4c_Ws_Ws'G')\alpha_d\;,\\
		k_6&=&1+2s_Ws'c'G'+s'^2s_W^2G'^2+2c_Ws_Ws'^2\alpha_a-c_W^2s'^2\alpha_b\nonumber\\
		&&-(4c_Ws'c'+4c_Ws_Ws'^2G')\alpha_c+(4s_Ws'c'+4s'^2s_W^2G')\alpha_d\;.
\label{ksDef}
\end{eqnarray}
Note that as long as we have a nonzero Stueckelberg coupling $G'$, the rotated kinetic matrix $\tilde{U}_0^TU_0^T\mathcal{K}_0U_0\tilde{U}_0$ is not diagonal, even if the kinetic mixing coefficients $\alpha_a$,$\alpha_b$,$\alpha_c$,$\alpha_d$ all vanish. For the special case $g''=\tilde{g}'=0$, the matrix elements reduce to $k_3=k_5=0$ and $k_6=1$.

With these results, we introduce the matrix $\tilde{U}_1$ to further diagonalize the rotated kinetic matrix $\tilde{U}_0^TU_0^T\mathcal{K}_0U_0\tilde{U}_0$
\begin{eqnarray}
\tilde{U}_1\!\equiv\tilde{U}_0^{-1}U_1\!
=\left(\begin{array}{ccc}
        l\cos(\theta_l\!-\!\theta') & 0 & r\sin(\theta_r\!-\!\theta')
        \\
        \xi & a & \eta
        \\
        -l\sin(\theta_l\!-\!\theta') & 0 & r\cos(\theta_r\!-\!\theta')
    \end{array}\right)\;,~~~~~\label{kineticMixing}
\end{eqnarray}
which changes the diagonal matrix $\mathrm{diag}(M^2,0,{M'}^2)$ to $\mathrm{diag}(M_Z^2,0,M_{Z'}^2)$ with
\begin{eqnarray}
M_Z^2&=&M^2l^2\left[\cos^2(\theta_l-\theta')+\frac{\cos(\theta_l-\theta')\sin(\theta_r-\theta')\sin(\theta_l-\theta')}{\cos(\theta_r-\theta')}\right]\;,\\
M_{Z'}^2&=&{M'}^2r^2\left[\cos^2(\theta_r-\theta')+\frac{\cos(\theta_r-\theta')\sin(\theta_r-\theta')\sin(\theta_l
-\theta')}{\cos(\theta_l-\theta')}\right]\;,
\end{eqnarray}
as long as we take
\begin{eqnarray}
\frac{\tan(\theta_l-\theta')}{\tan(\theta_r-\theta')}=\frac{M^2}{M'^2}\;.\label{thetarEq}
\end{eqnarray}
 i.e.
 \begin{eqnarray}
 \tilde{U}_1^T\tilde{U}_0^TU_0^T\mathcal{K}_0U_0\tilde{U}_0\tilde{U}_1
 =U_1^TU_0^T\mathcal{K}_0U_0U_1
=U^T\mathcal{K}_0U=-\frac{1}{4}\mathrm{diag}(1,1,1)\;.\label{kineticDiag}
 \end{eqnarray}
We see that the parameters in (\ref{kineticMixing})
 play the role of generating most general {\it kinetic} mixings. In particular, $\xi$ and $\eta$ originate from $Z-\gamma$ and $Z'-\gamma$ mixings respectively, while $l$, $r$, $\theta_l-\theta'$ and $\theta_r-\theta'$ originate from the most general $Z$ and $Z'$ redefinition and mixing which need four independent parameters (two from redefinition and the other two from kinetic mixings).

 The $\theta'$ appearing in (\ref{kineticMixing}) in the combinations of $\theta_l-\theta'$ and $\theta_r-\theta'$ is needed to subtract out $Z$-$Z'$ mass mixing from general $Z$-$\gamma$-$Z'$ mixings, leaving only the pure kinetic mixings.  If there are no kinetic mixings, then
\begin{eqnarray}
a=l=r=1\hspace*{2cm}G'=\xi=\eta=0\hspace*{2cm}\theta_l=\theta_r=\theta'\;.\label{ZZ'mixing}
\end{eqnarray}
By further requiring no $Z$-$Z'$ mass mixing by taking $\theta'=0$ in above result, we recover the SM tree diagram limit mentioned previously.

Using (\ref{kineticDiag}), we then find
\begin{eqnarray}
\frac{1}{a^2}=k_4\label{aEq}
\end{eqnarray}
which only rescales the photon field to normalized kinetic form. Equation (\ref{thetarEq}) gives one relation between the angle combinations $\theta_l-\theta'$ and $\theta_r-\theta'$,  (\ref{kineticDiag}) further fixes $\tan(\theta_l-\theta')$ through the following quadratic equation
\begin{eqnarray}
&&\Big\{\frac{k_2k_5}{k_4}-k_3\Big\}M^2M'^2\frac{\tan^2(\theta_l-\theta')}{M^4}\nonumber
	\\
	&&+\Big\{(k_1-\frac{k_2^2}{k_4})M'^2+(\frac{k_5^2}{k_4}-k_6)M^2\Big\}\frac{\tan(\theta_l-\theta')}{M^2}
	+\Big\{k_3-\frac{k_2k_5}{k_4}\Big\}=0\label{thetalEq}
\end{eqnarray}
There are two solutions from the above equation: one of these we choose so that it vanishes in the limit  $k_1=k_4=k_6=1,k_2=k_3=k_5=0$ for fixed $M^2$ and $M'^2$, the other nonzero solution corresponds to having the$Z$ mass vanish and $\gamma$ receiving a nonzero mass. Combining  the solution of (\ref{thetalEq}) with equation (\ref{thetarEq}), we obtain $\theta_l-\theta'$ and $\theta_r-\theta'$.  $r$ and $l$ can be determined by
\begin{eqnarray}
&&\hspace*{-1cm}\frac{1}{l^2}=\cos^2(\theta_l-\theta')\Big\{(k_6-\frac{k_5^2}{k_4})\tan^2(\theta_l-\theta')
	+2(\frac{k_2k_5}{k_4}-k_3)\tan(\theta_l-\theta')
	+k_1-\frac{k_2^2}{k_4}\Big\}
\label{lEq}~~\\
&&\hspace*{-1cm}\frac{1}{r^2}=\cos^2(\theta_r\!-\!\theta')\Big\{(k_1-\frac{k_2^2}{k_4})\tan^2(\theta_r-\theta')
	+2(k_3-\frac{k_2k_5}{k_4})\tan(\theta_r-\theta')
	+k_6-\frac{k_5^2}{k_4}\Big\}\label{rEq}
\end{eqnarray}
With $l$, $r$, $\theta_l-\theta'$ and $\theta_r-\theta'$, $\xi$ known, and $\eta$ are re-expressible
\begin{eqnarray}
\frac{\xi}{l}&=&\frac{k_5\sin(\theta_l-\theta')-k_2\cos(\theta_l-\theta')}{k_4}\label{xiEq}\\
\frac{\eta}{r}&=&-\frac{k_2\sin(\theta_r-\theta')+k_5\cos(\theta_r-\theta')}{k_4}\;.\label{etaEq}
\end{eqnarray}
As an example, we give the explicit result for the special case $g''=\tilde{g}=0$, (present situation is $0/0$ case, here in the limiting procedure we let $\tilde{g}$ approach  zero first, then take $g''$ to zero, because as we mentioned before $G'$ is small from purely phenomenological estimations.) where the above considerations program  gives result:
\begin{eqnarray}
&&\hspace*{-0.5cm}\theta_l=\theta_r=\theta'=G'=\eta=0\hspace{1cm}\frac{1}{a^2}=k_4\hspace{1cm}\frac{1}{l^2}=k_1\!-\frac{k_2^2}{k_4}\hspace{1cm}r=1\hspace{1cm}
\xi=-\frac{k_2l}{k_4}\\
&&\hspace{-0.5cm}M^2_Z\!=M^2l^2\hspace{3.5cm}M^2=\frac{1}{4}\bar{g}_Z^2f^2\hspace{3.2cm}M_{Z'}^2\!=M^{\prime 2}\!=0
\end{eqnarray}

Up to this stage, once we know the coefficients in mass-squared matrix $\mathcal{M}_0^2$ and kinetic matrix $\mathcal{K}_0$, i.e. $f$ and eight theoretical coefficients of EWCL $\bar{g}/\bar{g}', \bar{\tilde{g}}'/\bar{g}'', \bar{g}_Z/\bar{g}'', \bar{\beta_2},\alpha_a,  \alpha_b, \alpha_c, \alpha_d$, we can obtain the final phenomenological mixing parameters $\theta_W,\theta_r,\theta_l,G', \xi,\eta,l,r$, and intermediate mixing angle $\theta'$, photon normalization factor $a$.
In particular, the intermediate mass-squared ratio $M^2/M'^2$ is determined from (\ref{thetarEq}) and the physical mass ratio $M_Z/M_{Z'}$ can be expressed as
\begin{eqnarray}
\frac{M_Z}{M_{Z'}}=\frac{l}{r}\frac{\sin^{1/2}(2\theta_l-2\theta')}{\sin^{1/2}(2\theta_r-2\theta')}\;.\label{massratio}
\end{eqnarray}
This result offers a hope in predicting the $Z'$ mass in mixing parameters. Unfortunately,  the mixing parameters themselves are not easy to test. In the next section, we will discuss the experimental measurability of the mixing parameters. Here we would rather treat the above relation  as an additional constraint used in determining parameters for a given $Z-Z'$ mass ratio.

Phenomenologically, a more important question is, once we know the eight phenomenological mixing parameter $\theta_W,\theta_r,\theta_l,G',\xi,\eta,l,r$ from fitting the experiment data, how can we obtain the corresponding eight theoretical  coefficients $\bar{g}/\bar{g}',\bar{\tilde{g}}'/\bar{g}'',\bar{g_Z}/\bar{g}'', \bar{\beta_2}, \alpha_a, \alpha_b, \alpha_c, \alpha_d$.
Considering that the mixing parameter $G'=\bar{\tilde{g}}'/g''$ has already appeared in $\mathcal{M}_0^2$, i.e. it is both a theoretical coefficient and a phenomenological  parameter, the problem remaining is to fix the other seven coefficients $\bar{g}/\bar{g}', \bar{g}_Z/\bar{g}'', \bar{\beta_2}, \alpha_a,\alpha_b,\alpha_c,\alpha_d$ in  eight phenomenological parameters $\theta_W,\theta_r,\theta_l,G',\xi,\eta,l,r$.
Since the details of  computation are very complex, here we only outline the calculations: We choose seven equations
(\ref{WeinbergDef}), (\ref{thetarEq}), (\ref{thetalEq}), (\ref{lEq}) ,  (\ref{rEq}),  (\ref{xiEq}), and (\ref{etaEq}) for which  the auxiliary quantity $\theta'$ is further determined by (\ref{thetaprime}), $M^2/M'^2$ by (\ref{MMprime}), and $k_1,k_2,k_3,k_4,k_5,k_6$ by (\ref{k1}) to (\ref{ksDef}). Solving these equations, we can in principle express these theoretical coefficients in phenomenological parameters.

With the expressions of the EWCL coefficients of the phenomenological parameters, and with help of (\ref{thetaprime}) and (\ref{aEq}), the conventional $Z$-$Z'$ mass mixing angle $\theta'$, the ratio $M_{Z}/M_{Z'}$ and $a$ can all be expressed in the eight phenomenological mixing parameters.

The above procedure yields completely general results. To terms of order  $p^4$, we give explicit expressions for six phenomenological parameters $\theta_r,\theta_l,\xi,\eta,l,r$ in terms of
theoretical coefficients $\bar{g}'/\bar{g},\bar{\tilde{g}}'/\bar{g}'',\bar{g}_Z/\bar{g}'', \bar{\beta}_2, \alpha_a, \alpha_b, \alpha_c, \alpha_d$:
\begin{eqnarray}
\theta_r&\simeq&\theta'+\frac{4s_W\bar{g}''^2}{\Delta_g}G'+\frac{s_W(5\bar{g}_Z^2+12\bar{g}''^2)}{\Delta_g}G'\theta'^2-\frac{4c_W(-2\bar{g}_Z^2s_W^2+\Delta_g)\bar{g}''^2}{\Delta_g^2}G'\alpha_a
	\nonumber\\
	&&-\frac{4s_Wc_W^2\bar{g}_Z^2\bar{g}''^2}{\Delta_g^2}G'\alpha_b
	-\frac{8c_W\bar{g}''^2}{\Delta_g}\alpha_c
	+\frac{8s_W\bar{g}''^2}{\Delta_g}\alpha_d
\label{thetar}\\
\theta_l&\simeq&\theta'
	+\frac{s_W\bar{g}_Z^2}{\Delta_g}G'+\frac{s_W(3\bar{g}_Z^2+20\bar{g}''^2)}{\Delta_g}G'\theta'^2
	-\frac{c_W\bar{g}_Z^2(-2\bar{g}_Z^2s_W^2+\Delta_g)}{\Delta_g^2}G'\alpha_a\nonumber
	\\
	&&-\frac{s_Wc_W^2\bar{g}_Z^4}{\Delta_g^2}G'\alpha_b
	-\frac{2c_W\bar{g}_Z^2}{\Delta_g}\alpha_c
	+\frac{2s_W\bar{g}_Z^2}{\Delta_g}\alpha_d
\label{thetal}\\
r&\simeq&1-s_WG'\theta'+\frac{2c_Ws_W(\bar{g}_Z^2+4\bar{g}''^2)}{\Delta_g}G'\alpha_c+\frac{2(c_W^2\bar{g}_Z^2+4(c_W^2-2)\bar{g}''^2)}{\Delta_g}G'\alpha_d\;,
\\
l&\simeq&1+s_WG'\theta'-s_Wc_W\alpha_a+\frac{c_W^2}{2}\alpha_b
	-\frac{2c_Ws_W(\bar{g}_Z^2+4\bar{g}''^2)}{\Delta_g}G'\alpha_c
	+\frac{2s_W^2(\bar{g}_Z^2+4\bar{g}''^2)}{\Delta_g}G'\alpha_d\;,~~~~\\
\xi&\simeq&-c_WG'\theta'+(2c_W^2-1)\alpha_a+c_Ws_W\alpha_b+\frac{8(2c_W^2-1)\bar{g}''^2}{\Delta_g}G_3\alpha_c
	-\frac{16c_Ws_W\bar{g}''^2}{\Delta_g}G'\alpha_d\;,
\\
\eta&\simeq&c_WG'-\frac{c_W}{2}G'\theta'^2+\frac{2s_W(c_W^2\bar{g}_Z^2-2\bar{g}''^2)}{\Delta_g}G'\alpha_a
	+\frac{\bar{g}_Z^2c_Ws_W^2}{\Delta_g}G'\alpha_b+2s_W\alpha_c+2c_W\alpha_d\;.
\end{eqnarray}
Here, $\theta'\simeq -2\bar{g}_Z\bar{g}''\bar{\beta}_2/\Delta_g$, $\theta_W=\arctan \bar{g}'/\bar{g}$ and $G'=\bar{\tilde{g}}'/\bar{g}''$. Moreover, we obtain,
\begin{eqnarray}
&&a\simeq 1+c_Ws_W\alpha_a+\frac{s_W^2}{2}\alpha_b-2c_Ws_WG'\alpha_c-2c_W^2G'\alpha_d\;,\\
&&\theta'\simeq\frac{\bar{g}_Z^2\theta_r-4\bar{g}''^2\theta_l}{\Delta_g}\;.\label{thetap}
\end{eqnarray}
Note that since (\ref{ZZ'mixing}) tells us that if there is no kinetic mixings, $\theta_l=\theta_r=\theta'$, then the differences
$\theta_l-\theta'$ and $\theta_r-\theta'$ reflect the effects caused by kinetic mixings. Substituting (\ref{thetar}) and (\ref{thetal}) into (\ref{thetarEq}), we find the result for $M^2/M'^2$ which just matches the results that we obtained from (\ref{MMprime}). Although our result here already includes in all possible mixings cases, pure $Z$-$Z'$ mass mixing is worth a special discussion: we find that the limit $G'=\alpha_c=\alpha_d=0$ can not be taken at the very beginning, since this will lead to $\theta_r=\theta_l=\theta'$ from (\ref{thetar}) to (\ref{thetal}) and then limit problems $0/0$ in (\ref{thetarEq}) for $M^2/M'^2$. To obtain the correct result, we need first to maintain $G'$ and $\alpha_c$, $\alpha_d$ with nonzero values through to completion of the computation of the ratio $M^2/M'^2$, then taking its vanishing limit. This is an interesting new phenomena, i.e. nonzero $G'$ and $\alpha_c$, $\alpha_d$ extensions make that $M^2/M'^2$ can be expressed in mixing parameters. In contrast with the pure $Z$-$Z'$ mass mixing case that from (\ref{MMprime}) we find that just the mixing angle $\theta'$ can not fully fix the value of $M^2/M'^2$ as we are left with  $\bar{\beta}_2$ degrees of freedom remaining.

\section{ Measurability of the parameters and relevant EWCL coefficients}

Compared with the coefficients in EWCL, our eight parameters $\theta_W$, $G'$, $\xi$, $\eta$, $\theta_l$, $\theta_r$, $r$, $l$ are more close to experimental data and more easily determined by experiment. Once these are known, the relevant EWCL coefficients can be further determined by establishing relations between these parameters and the EWCL coefficients. In this section, we begin by discussing how these parameter values can be fixed in principle from experiment, and then construct the relations among the EWCL coefficients and parameters.

Experimentally, with the exception of $SU(2)_L$ coupling $g$ which can be determined from charged currents, the main means to determine the mixing parameters is by testing the structure of the electro-magnetic and neutral currents. The corresponding Lagrangian is $gW^3_{\mu}J^{3,\mu}+g'B_{\mu}J_Y^\mu+g''X_{\mu}J_X^\mu$, where $J^{3,\mu}$ is the third component of the conventional weak isospin current, $J_Y^\mu$ is the hypercharge current, and $J_X^\mu$ is the current coupled to the extra $X_\mu$ boson. The Lagrangian of the electro-magnetic and neutral currents that couple to the physical bosons $\gamma,Z,Z'$ becomes $eJ_{\mathrm{em}}^{\mu}A_\mu+g_ZJ_Z^{\mu}Z_\mu+g''J_{Z'}^{\mu}Z'_\mu$. With the help of (\ref{Uintr}), we can read off
\begin{eqnarray}
&&\hspace*{-0.5cm}eJ_{\mathrm{em}}^{\mu}=gU_{1,2}J^{3,\mu}+g'U_{2,2}J_Y^\mu+g''U_{3,2}J_X^\mu=gs_Wa[J^{3,\mu}+J_Y^\mu]+g''U_{3,2}J_X^\mu\label{Jem}\\
&&\hspace*{-0.5cm}g_ZJ_Z^{\mu}=gU_{1,1}J^{3,\mu}+g'U_{2,1}J_Y^\mu+g''U_{3,1}J_X^\mu\nonumber\\
&&\hspace{0.5cm}=g[(s_W\xi+c_Wc_ll)J^{3,\mu}+(s_W\xi-s_Wc_ll\tan\theta_W)J_Y^\mu]+g''U_{3,1}J_X^\mu\label{JZ}\\
&&\hspace*{-0.5cm}g''J_{Z'}^{\mu}=gU_{1,3}J^{3,\mu}+g'U_{2,3}J_Y^\mu+g''U_{3,3}J_X^\mu\label{JZprime}\;,
\end{eqnarray}
with $U_{i,j}$ a general matrix element of mixing matrix $U$, and we have used the result $gU_{1,2}=g'U_{2,2}$ combined (\ref{Udef}) and (\ref{WeinbergDef}). In principle, once experiments finally fix the coefficients $U_{i,j}$, then from (\ref{Udef}), we can determine all  eight parameters $\theta_W$, $G'$, $\xi$, $\eta$, $\theta_l$, $\theta_r$, $r$, $l$. Considering that  $Z'$ has not been discovered as yet in current experiments, we divide the present experimental  measurability of the parameters into two stages:
\begin{enumerate}
 \item Suppose we can measure $eJ_{\mathrm{em}}^{\mu}$ and $g_ZJ_Z^{\mu}$ experimentally but not know what $J_{Z'}^\mu$ and $J_X^\mu$ are. This is the present SM situation as it stands and is independent of details of the  $Z'$ model. Then (\ref{Jem}) implies that we can determine $gs_Wa$ and the electro-magnetic coupling $e$ now must be identified as $e=gs_Wa$. Compared with conventional relation in SM, we find that an extra correction factor $a$ appears in the relation.  Considering that $e$ and $g$ can be measured from electro-magnetic and charge currents respectively, we can then derive $s_Wa$. Further, from (\ref{JZ}), we find $g(s_W\xi+c_Wc_ll)$ and $g(s_W\xi-s_Wc_ll\tan\theta_W)$. Then, in this first stage, combined with known $g$, we can obtain four combinations of the eight parameters: $g$, $s_Wa$, $s_W\xi+c_Wc_ll$ and $s_W\xi-s_Wc_ll\tan\theta_W$.
\item Suppose in addition to $eJ_{\mathrm{em}}^{\mu}$ and $g_ZJ_Z^{\mu}$, we also know $J_X^\mu$. This can be realized if we have a prior $U(1)'$ charges for the SM fermions which is $Z'$ model-dependent. Then from (\ref{Jem}) and (\ref{Udef}), $g''U_{3,2}=g''(s_W\eta+c_Ws_rr)$ is obtainable; from (\ref{JZ}) and (\ref{Udef}), $g''U_{3,1}=g''(c_W\eta-s_Ws_rr)$ is calculable. We find at this second stage, we can obtain a further two combinations of the eight parameters.
\end{enumerate}
Therefore, before needing to measure $g''J_{Z'}^{\mu}$, the above two stages already enable us evaluate seven of the eight parameters. Using (\ref{massratio}), the remaining unknown parameter can be determined once we assume a $Z-Z'$ mass ratio. Thus, even without the knowledge of $g''J_{Z'}^{\mu}$, and as long as the $Z-Z'$ mass ratio is fixed, we can now measure all eight phenomenological parameters.

In consequence, we can express the EWCL coefficients in these parameters. Up to order $p^4$, the theoretical coefficients $\bar{g}_Z/\bar{g}'', \bar{\beta}_2, \alpha_a, \alpha_b, \alpha_c, \alpha_d$ in phenomenological parameters $\theta_W, \theta_r,\theta_l,G', \xi,\eta,l,r$ can be written as
\begin{eqnarray}
\frac{\bar{g}_Z}{\bar{g}''}&\simeq&\frac{2(\theta_l-\theta')}{\theta_r-\theta'}\;,\label{gZgpp}\\
\bar{\beta}_2&\simeq&-\frac{\bar{g}_Z^2\theta_r-4\bar{g}''^2\theta_l}{2\bar{g}_Z\bar{g}''}\;,\\
\alpha_a&=&-\frac{1}{4s_Wc_W\bar{g}''^2\Delta_g}\Big\{
	s_W(\bar{g}_Z^2s_W^2+2(c_W^2-2)\bar{g}''^2)\Delta_gG'\theta'\nonumber
	\\
	&&+8s_W^2\bar{g}''^2\Delta_g(l-1)+(\bar{g}_Z^2s_W^2+(4-2c_W^2)\bar{g}''^2)\Delta_g(r-1)\nonumber
	\\
	&&-4c_Ws_W\bar{g}''^2\Delta_g \xi+c_W(-\bar{g}_Z^4s_W^2-2\bar{g}''^2c_W^2\Delta_g)G'\eta\Big\}
\\
\alpha_b&=&-\frac{1}{4c_W^2\bar{g}''^2\Delta_g}\Big\{
	s_W(\Delta_g-2c_W^2\bar{g}_Z^2+4c_W^2\bar{g}''^2)\Delta_gG'\theta'
	\nonumber\\
	&&+8\bar{g}''^2(1-2c_W^2)\Delta_g (l-1)	
   +((1-2c_W^2)\bar{g}_Z^2+4\bar{g}''^2s_W^2)\Delta_g(r-1)
	\nonumber\\
	&&-8s_Wc_W\bar{g}''^2\Delta_g \xi
	+c_W(-\bar{g}_Z^4s_W^2-16\bar{g}''^4s_W^2+\bar{g}_Z^2c_W^2\Delta_g)G'\eta
	\Big\}
\\
\alpha_c&=&\frac{1}{8s_Wc_W\bar{g}''^2\Delta_g}\Big\{
	s_Wc^2_W\Delta_g^2\theta'+s^2_Wc^2_W(5\bar{g}_Z^2+14s_W\bar{g}''^2)\Delta_gG'\theta'^2\nonumber
	\\
	&&-c_W^2s_W\Delta_g^2\theta_r
	-8s_W^2\bar{g}''^2(-\bar{g}_Z^2s_W^2+4\bar{g}''^2)G'(l-1)\nonumber
	\\
	&&+(s_W^4\bar{g}_Z^4-16\bar{g}''^4+2\bar{g}''^2c_W^4\bar{g}_Z^2+8\bar{g}''^4c_W^2)G'(r-1)\nonumber
	\\
	&&+4s_Wc_W\bar{g}''^2(\bar{g}_Z^2(c_W^2-2)+4\bar{g}''^2)G'\xi
	+4s_W^2c_W\bar{g}''^2\Delta_g\eta\Big\}
\\
\alpha_d&=&-\frac{1}{8\bar{g}''^2\Delta_g}\Big\{
	s_W\Delta_g^2\theta'
	+4\bar{g}''^2\Delta_gG'
	+(5\bar{g}_Z^2s_W^2+(12-14c_W^2)\bar{g}''^2)\Delta_gG'\theta'^2
	\\
	&&-s_W\Delta_g^2\theta_r
	-8s_W^2\bar{g}_Z^2\bar{g}''^2G'(l-1)
	+\bar{g}_Z^2(-\bar{g}_Z^2s_W^2+(-4+2c_W^2)\bar{g}''^2)G'(r-1)
	\nonumber\\
	&&+4s_Wc_W\bar{g}_Z^2\bar{g}''^2G'\xi
	-4c_W\bar{g}''^2\Delta_g\eta
	\Big\}
\label{alphad}
\end{eqnarray}
Where $\theta'$ is given by (\ref{thetap}) and $\bar{g}_Z/\bar{g}''$ is given by (\ref{gZgpp}). The remaining two theoretical coefficients
  $\bar{g}'/\bar{g}$ and $\bar{\tilde{g}}'/\bar{g}''$, which are already determined in (\ref{WeinbergDef}) and (\ref{Gp}) respectively, are not displayed with the above formulae. Substituting the results back into (\ref{aEq}) and combining with (\ref{k1}), we further obtain
  \begin{eqnarray}
  a&=&1-\frac{1}{8c_W^2\bar{g}''^2\Delta_g}\Big\{
	s_W(s_W^2\bar{g}_Z^2-4\bar{g}''^2)\Delta_gG'\theta'
	+8s_W^2\bar{g}''^2\Delta_g (l-1)\label{a}\\
	&&
	+(\bar{g}_Z^2s_W^2+4\bar{g}''^2)\Delta_g (r-1)-8s_Wc_W\bar{g}''^2\Delta_g \xi
	-c_W(\bar{g}_Z^4s_W^2-4\bar{g}_Z^2\bar{g}''^2c_W^2+16\bar{g}''^4)G'\eta
	\Big\}\nonumber
\end{eqnarray}
 The results (\ref{thetap}) to (\ref{a}) indicate that once we known the eight phenomenological parameters $\theta_W, G',\xi,\eta,\theta_l,\theta_r,r,l$, the conventional $Z$-$Z'$ mixing angle $\theta'$, then the general $Z$-$\gamma$-$Z'$ mixing coefficients $\bar{g}/\bar{g}', \bar{g}_Z/\bar{g}'', G', \bar{\beta}_2, \alpha_a, \alpha_b, \alpha_c, \alpha_d$, and $a$ are fixed, where the $a$ parameter although appears in phenomenological role, as discussed earlier, it is derivable from the other eight parameters through (\ref{a}).

\section{Summary}

To summarize our results, based on the extended electroweak chiral Lagrangian previously proposed by us, we have found that there are eight independent degrees of freedoms to describe the most general $Z$-$\gamma$-$Z'$ mixings that correspond to the eight independent theoretical coefficients  $\bar{g}/\bar{g}',\bar{\tilde{g}}'/\bar{g}'',\bar{g}_Z/\bar{g}''$, $\bar{\beta}_2$, $\alpha_a$, $\alpha_b$, $\alpha_c$, $\alpha_d$ in our electroweak chiral Lagrangian.  For convenience in phenomenological analysis, we have proposed a new general parameterization involving these eight parameters that describe the $Z$-$\gamma$-$Z'$ mixings, which include the conventional Weinberg angle $\theta_W$ and a Stueckelberg-type coupling $G'$.  Combined with the conventional $Z$-$Z'$ mass mixing parameter $\theta'$, we find that parameters $\xi$, $\eta$, $\theta_l$-$\theta'$, $\theta_r$-$\theta'$, $r$, $l$ reflect the general kinetic mixings
among the $Z$-$\gamma$-$Z'$. With this  parameterization, $\theta_W$, $G'$, $\xi$, $\eta$, $\theta_l$, $\theta_r$, $r$, $l$, we can fully determine the $Z$-$Z'$ mass mixing angle $\theta'$ and the mass ratio $M_{Z}/M_{Z'}$. Experimentally,  with the knowledge of charge currents, neutral currents and the current for extra gauge boson $X_\mu$, combined with  mass ratio $M_{Z}/M_{Z'}$, we can in principle measure all  eight parameters.

\section*{Acknowledgments}

This work was  supported by National  Science Foundation of China
(NSFC) under Grant No. 10875065 and No.10947152.

\end{document}